\documentclass[prd,aps,twocolumn,preprintnumbers,amsmath,amssymb,nofootinbib,superscriptaddress,notitlepage]{revtex4-1}

\usepackage{epsfig}
\usepackage[utf8]{inputenc}
\usepackage{color}
\usepackage[colorlinks=true,
linkcolor=blue,
breaklinks=true,
urlcolor=blue,
citecolor=blue]{hyperref}

\usepackage{graphicx}
\usepackage{bm}

\newcommand{\be}{\begin{equation}}
\newcommand{\ee}{\end{equation}}
\newcommand{\bea}{\begin{eqnarray}}
\newcommand{\eea}{\end{eqnarray}}
\newcommand{\ba}{\begin{array}}
\newcommand{\ea}{\end{array}}
\newcommand{\bi}{\begin{itemize}}
\newcommand{\ei}{\end{itemize}}

\newcommand{\ucas}{\affiliation{University of Chinese Academy of Sciences, Beijing 100049, China}}

\newcommand{\imp}{\affiliation{Institute of Modern Physics, Chinese Academy of Sciences, Lanzhou 730000, China}}

\newcommand{\shandong}{\affiliation{Institute of Frontier and Interdisciplinary Science and Key Laboratory of Particle Physics and Particle Irradiation (MOE), Shandong University, Qingdao, Shandong 266237, China}}

\begin{document}

\title{Extrapolating Exclusive Charmonium Photoproduction to the Forward Regime Using Gaussian Process}

\author{Xue Wang}\email{xuewang@mail.sdu.edu.cn}
\shandong

\author{Xu Cao}\email{caoxu@impcas.ac.cn}
%\thanks{\textcolor{red}{Corresponding author.}}
\imp
\ucas
%\csr

\author{Yu Lu}\email{ylu@ucas.ac.cn}
\thanks{(Corresponding author)}
\ucas

\author{Weizhi Xiong}\email{xiongw@mail.sdu.edu.cn}
\shandong

\date{\today}

\begin{abstract}

Accessing the forward kinematic regime of exclusive vector meson photoproduction is essential for probing the structure of the proton, 
yet this region is often experimentally inaccessible. 
We present a novel extrapolation of near-threshold $\gamma p \to J/\psi p$ differential cross section data to the forward regime $d\sigma/dt|_{t=0}$ by applying a non-parametric Gaussian Process (GP) regression. 
By employing a two-dimensional GP dependent on both the squared momentum transfer $t$ and the center-of-mass energy $W$, 
we perform a coherent analysis of world data and provide a robust quantification of the extrapolation uncertainties. 
Furthermore, we apply the GP in the determination of the proton mass radius from the $J/\psi$ photoproduction data.
We find that the extracted radius is subject to large systematic uncertainties, 
highlighting the challenge in precisely constraining this quantity with current experimental precision. 
This work demonstrates that GP is a powerful and systematic tool for phenomenological studies in hadron physics, 
particularly for extrapolations into unmeasured kinematic domains.

\end{abstract}

\maketitle

\section{Introduction} \label{sec:intro}

\begin{figure}[htbp]
  \centering
    \includegraphics[width=0.8\linewidth]{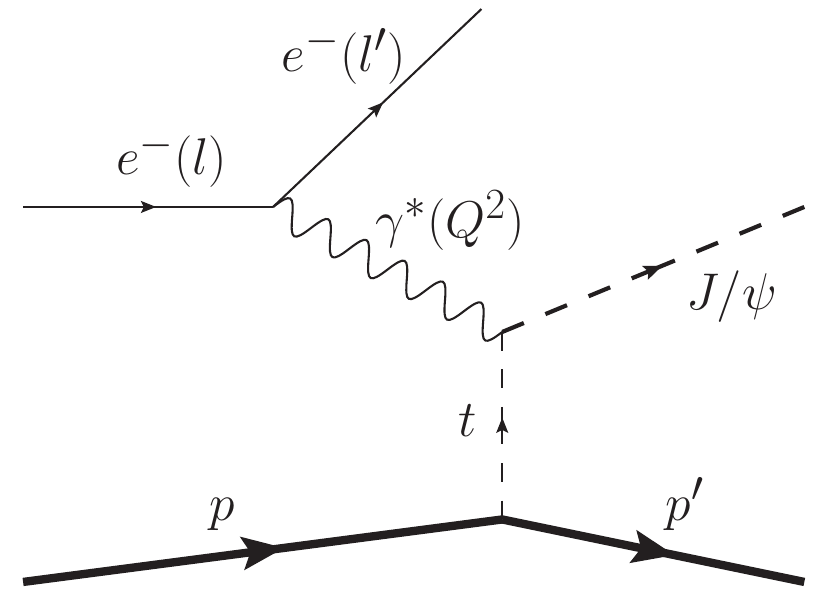}
  \caption{Leading-order $t$-channel diagram for the electro-production process $e^-p \to e^-p J/\psi$. 
  The relevant kinematic variables are labeled besides the lines with virtuality of photon $Q^2 = -q^2 = -(l'-l)^2$ and transfer momentum squared $t = (p' - p)^2$. 
  The invariant mass of the photon-proton system is defined as $W = \sqrt{(q+p)^2}$ and the total c.m energy as $s=(l + p)^2$.
  }
  \label{fig:feynman}
\end{figure}

The exclusive photoproduction of charmonium off a proton, $\gamma p \to J/\psi p$ (Fig.~\ref{fig:feynman}), 
is a powerful probe of the proton's internal structure. 
Near the production threshold, the scattering amplitude in the forward regime
($t \to 0$, where $t$ is the squared momentum transfer) encodes essential information about the gluon Generalized Parton Distribution (GPD) \cite{Diehl:2003ny,Belitsky:2005qn,Guo:2021ibg}, 
the charmonium-proton interaction \cite{Gryniuk:2016mpk}, 
and the search for exotic pentaquark states \cite{Cao:2019kst,JointPhysicsAnalysisCenter:2023qgg}. 
However, this forward, near-threshold region is experimentally challenging due to kinematic constraints and detector acceptance, necessitating a robust extrapolation from existing data.

Current analyses typically rely on parametric phenomenological models, 
such as assuming a simple exponential or multi-pole form for the $t$-dependence of the differential cross section \cite{Guo:2023qgu,Sun:2021pyw,Sun:2021gmi,Kharzeev:2021qkd}. 
While practical, this approach introduces significant model dependence, 
and the systematic uncertainties arising from the rigidity of these functional forms are often not fully explored. 
This is particularly critical for extracting the forward slope of the cross section, 
a quantity related to the proton's gravitational form factors and mass radius \cite{Guo:2025jiz}.

To address these limitations, we employ a non-parametric, data-driven approach using Gaussian Process (GP) regression. 
A GP defines a probability distribution over a space of functions, 
specified by a kernel whose hyperparameters are learned directly from the data \cite{rasmussen:2006gau}. 
This framework avoids pre-defined assumptions about the functional form of the underlying physics, 
thereby mitigating model bias. 
A key feature of GP is its inherent ability to provide both a mean prediction and a principled, data-driven uncertainty estimate.

The use of GPs and other Bayesian methods is becoming increasingly prevalent in hadron and nuclear physics. 
They have been applied to quantify interpolation uncertainties in theoretical models \cite{Ferguson:2025ddu,Jaiswal:2025hyp}, 
assess hadronic contributions to the muon's anomalous magnetic moment \cite{Fowlie:2023cta}, 
reconstruct parton distributions from lattice QCD \cite{Candido:2024hjt,Dutrieux:2025jed,Chen:2025cxr}, 
and estimate experimental backgrounds at the LHC \cite{Barr:2025lba}.

In this work, we extend the application of GPs to the extrapolation problem in exclusive $J/\psi$ photoproduction. 
By modeling the differential cross section as a two-dimensional function of $t$ and the center-of-mass energy $W$, 
we perform a model-independent extrapolation to the $t \to 0$ limit. 
Our goal is to provide a robust determination of the forward cross section and its slope, 
accompanied by a systematic quantification of the extrapolation uncertainty.

\section{The Gaussian Process} \label{sec:gp}

A GP essentially defines a distribution over functions,
such that the function values at any finite set of points follow a multivariate Gaussian distribution. 
Analogous to a multivariate Gaussian, 
which is fully specified by a mean vector $\bm{\mu}$ and a covariance matrix $\bm{\Sigma}$, 
a GP is uniquely determined by its mean function $\mu(\bm{x})$ and a covariance (or kernel) function $k(\bm{x}, \bm{x}')$.
This can be compactly expressed as
\bea
f(\bm{x}) \sim \mathcal{GP}(\mu(\bm{x}), k(\bm{x}, \bm{x}'))
\eea
In our analysis, the input $\bm{x}$ is a two-dimensional vector representing the kinematic variables, i.e., $\bm{x}=(W,t)$, 
and the function $f(x)$ corresponds to the logarithm of the differential cross-section.
The properties of the function space are primarily determined by the kernel $k(\bm{x}, \bm{x}')$,
which must be a positive semi-definite function.
The flexibility in choosing the kernel is a key source of the power of GP.

The power of a GP lies in the choice of its kernel function, 
which encodes prior assumptions about the function being modeled, such as smoothness. 
We employ the widely-used Squared Exponential (SE) kernel, which corresponds to infinitely differentiable functions:
\bea
k_{SE}(\bm{x}, \bm{x}') = \sigma_f^2 \exp\left(-\frac{1}{2} \frac{(x - x')^2}{l^2}\right) \label{eq:se-kernel-0}
\eea
Here, the hyperparameter $\sigma_f^2$ is the signal variance, controlling the overall amplitude of the function, 
and $l$ is the length-scale, which determines the smoothness. 
A larger length-scale implies that function values at distant points are more strongly correlated. 
We physically anticipate the differential cross-section to be a smooth function of both $t$ and $W$. 
For our two-dimensional input space $\bm{x}=(W,t)$, we use an anisotropic SE kernel:
\bea
K(\bm{x}_a, \bm{x}_b) = \sigma^2 \exp\left(-\frac{1}{2} \sum_{i=1}^{2} \frac{(x_{a,i} - x_{b,i})^2}{l_i^2}\right)\label{eq:se-kernel}
\eea
where $l_1$ and $l_2$ are the characteristic length-scales for the $W$ and $t$ dimensions, respectively, 
allowing for different degrees of correlation along each axis.

A complete GP specification also requires a mean function $\mu(\bm{x})$. 
A common choice is a zero-mean function, 
$\mu(\bm{x})=0$, which is reasonable when there is no prior knowledge about the baseline of the function. 
However, this is unsuitable for the differential cross section $d\sigma/dt$, 
which is strictly positive and spans several orders of magnitude (see Fig.~\ref{fig:dsigma_dt_t}).
To address this, we model the logarithm of the cross section, $f(\bm{x}) = \ln(d\sigma/dt)$. 
Furthermore, to better match the zero-mean prior, 
we pre-process the training data by subtracting its mean value. 
The GP is therefore trained on the residuals, $y_i - \bar{y}$, 
where $y_i = \ln(d\sigma/dt)_i$ and $\bar{y}$ is the sample mean of the training data.

With the optimized hyperparameters, the predictive distribution for the function value $f_* = f(\bm{x}_*)$ at a new point $\bm{x}_*$ is also Gaussian, with mean $\mu(f_*)$ and variance $\sigma^2(f_*)$ given by:
\begin{align}
\mu(f_*) &= \bm{k}_*^T (\bm{K} + \Sigma_{\mathrm{noise}})^{-1} \bm{y}, \label{eq:f_mu}\\
\sigma^2 (f_*) &= k(\bm{x}_*, \bm{x}_*) - \bm{k}_*^T (\bm{K} + \Sigma_{\mathrm{noise}})^{-1} \bm{k}_*, \label{eq:f_cov}
\end{align}
where $\bm{k}_* = [k(\bm{x}_*, \bm{x}_1), \ldots, k(\bm{x}_*, \bm{x}_n)]^T$ is the vector of covariances between the test point $\bm{x}_*$ and the training points.

The hyperparameters $\bm{\theta} := \{\sigma_f^2, l_1, l_2\}$ are optimized by maximizing the log marginal likelihood of the training data $\bm{y}$:
\begin{align}
  \log P(\bm{y}|X,\bm{\theta} ) &= -\frac{1}{2} \bm{y}^T (\bm{K} + \Sigma_{\mathrm{noise}})^{-1} \bm{y} \\ \nonumber 
  &\quad - \frac{1}{2} \ln |\bm{K} + \Sigma_{\mathrm{noise}}| - \frac{n}{2} \ln (2\pi),\\ 
\Sigma_{\mathrm{noise}} &= \mathrm{diag}(\sigma_1^2, \sigma_2^2, \ldots, \sigma_n^2)
\end{align}
where $\bm{y}$ is the vector of $n$ training points, 
$\bm{K}$ is the $n \times n$ covariance matrix with elements $K_{ab} = K(\bm{x}_a, \bm{x}_b)$, 
and $\Sigma_{\mathrm{noise}} = \mathrm{diag}(\sigma_1^2, \ldots, \sigma_n^2)$ is the noise covariance matrix, 
which accounts for the known experimental uncertainties $\sigma_i$.

With the optimized hyperparameters, 
the predictive distribution for the function value $f_* = f(\bm{x}_*)$ at a new point $\bm{x}_*$ is also Gaussian, 
with mean $\mu(f_*)$ and variance $\sigma^2(f_*)$ given by:
\begin{align}
\mu(f(\bm{x}_*)) &= \bm{k}_*^T (\bm{K} + \Sigma_{\mathrm{noise}})^{-1} \bm{y}, \label{eq:f_mu}\\
\sigma^2 (f(\bm{x}_*)) &= k(\bm{x}_*, \bm{x}_*) - \bm{k}_*^T (\bm{K} + \Sigma_{\mathrm{noise}})^{-1} \bm{k}_* \label{eq:f_cov}
\end{align}
where $\bm{k}_* = [k(\bm{x}_*, \bm{x}_1), \ldots, k(\bm{x}_*, \bm{x}_n)]^T$ is the vector of covariances between the new point $\bm{x}_*$ and the training points $\bm{x}_i$.

To extract the forward slope parameter, 
$B(W) \propto \partial f / \partial t |_{t=0}$, 
we also require predictions for the derivative of the function $f$. 
This is possible because differentiation is a linear operator, 
and a GP transformed by a linear operator remains a GP. 
The required covariance functions involving the derivative with respect to a dimension $j$, $f'_{j}(\bm{x}) \equiv \partial f(\bm{x}) / \partial x_j$, 
can be obtained by differentiating the primary kernel $K(\bm{x}_a, \bm{x}_b)$:
\begin{align}
\mathrm{Cov}(f'_{j}(\bm{x}_a), f(\bm{x}_b)) &= \frac{\partial}{\partial x_{a,j}} K(\bm{x}_a, \bm{x}_b) \\ 
&= K(\bm{x}_a, \bm{x}_b) \frac{-(x_{a,j}-x_{b,j})}{l_j^2}, \\
\mathrm{Cov}(f'_{j}(\bm{x}_a), f'_{j}(\bm{x}_b)) &= \frac{\partial^2}{\partial x_{a,j} \partial x_{b,j}} K(\bm{x}_a, \bm{x}_b) \label{eq:se-fpp}\\
&= K(\bm{x}_a, \bm{x}_b) \left[\frac{1}{l_j^2}-\frac{(x_{a,j}-x_{b,j})^2}{l_j^4}\right].
\end{align}
The predictive mean and variance for the derivative $f'_* = f'_j(\bm{x}_*)$ are then given by:
\begin{align}
\mu(f'_*) &= (\partial_{j*}\bm{k}_*)^T (\bm{K} + \Sigma_{\mathrm{noise}})^{-1} \bm{y},\\
\sigma^2 (f'_*) &= k_{\partial_j,\partial_j}(\bm{x}_*, \bm{x}_*) - (\partial_{j*}\bm{k}_*)^T (\bm{K} + \Sigma_{\mathrm{noise}})^{-1} (\partial_{j*}\bm{k}_*),
\end{align}
where $\partial_{j*}\bm{k}_*$ is the vector of covariances $\mathrm{Cov}(f'_{j}(\bm{x}_*), f(\bm{x}_i))$, 
and $k_{\partial_j,\partial_j}(\bm{x}_*, \bm{x}_*)$ is the prior variance $\mathrm{Cov}(f'_{j}(\bm{x}_*), f'_{j}(\bm{x}_*))$ evaluated using Eq.~\eqref{eq:se-fpp}.

\section{Results and Discussion}

The squared momentum transfer $t$ for the reaction $\gamma p \to J/\psi p$ is kinematically constrained to the range $t_1 \le t \le t_0 < 0$ \cite{Cao:2023rhu}, 
where
\bea \label{eq:trange}
t_{0,1}(W, Q^2) &=& \frac{1}{2 W^2} \Big( (Q^2+M_N^2)(M_V^2-M_N^2) \nonumber \\
&& \pm \sqrt{\lambda(W^2,-Q^2,M_N^2) \lambda(W^2,M_V^2,M_N^2)} \Big)\nonumber \\
&& - \frac{1}{2}(W^2+Q^2-M_V^2-2 M_N^2).
\eea
Here, 
$M_N$ and $M_V$ are the nucleon and $J/\psi$ masses, respectively, 
and $\lambda(x,y,z) = x^2 + y^2 + z^2 -2xy - 2yz -2zx$ is the K\"{a}ll\'{e}n function.
Since the physical region does not extend to $t=0$, 
accessing the forward regime requires an extrapolation from $t = t_0$. 
Furthermore, near the threshold energy $W_{th} = M_V + M_N$, 
the two-body phase space suppresses the cross section, 
limiting the statistical precision of measurements \cite{Cao:2023rhu,Wang:2023thy}.

\begin{figure}
    \centering
    \includegraphics[width=1.0\linewidth]{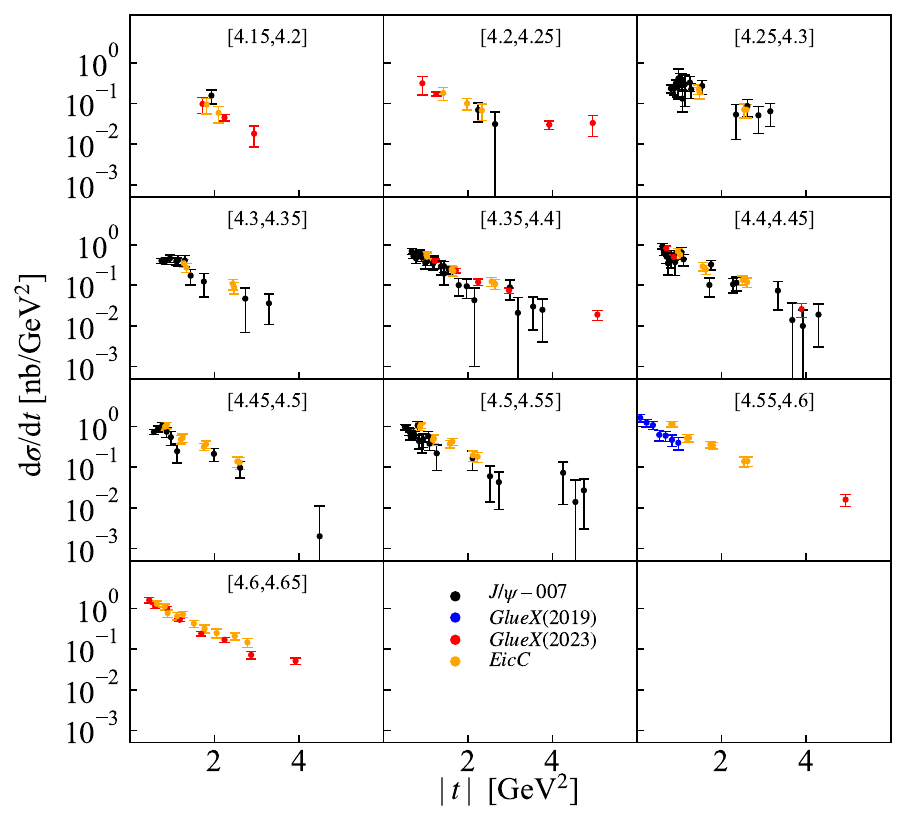}
    \caption{Comparison of data of differential cross sections $d\sigma/dt$ as a function of $|t|$ for variant $W$ bins labeled in square brackets.
   The experimental data are from GlueX(2019) \cite{GlueX:2019mkq} and GlueX(2023) \cite{GlueX:2023pev} and $J/\psi$-007 \cite{Duran:2022xag}. The pseudo-data at EicC is generated for quasi-real photon production ($Q^2 = 1.0$ GeV$^2$) \cite{Wang:2023thy}.}
    \label{fig:dsigma_dt_t}
\end{figure}

Our analysis utilizes existing near-threshold data from JLab ($W < 4.65$ GeV) and pseudo-data for a future Electron-ion collider in China (EicC). 
To validate our methodology, we generated the EicC pseudo-data by simulating quasi-real photoproduction ($Q^2 = 1.0~\mathrm{GeV}^2$) based on the JLab measurements, 
folding in the projected detector acceptance and resolution for an integrated luminosity of 50 fb$^{-1}$ \cite{Wang:2023thy}. 
The real and pseudo-data are shown in Fig.~\ref{fig:dsigma_dt_t}.

Initially, we perform a one-dimensional GP regression on the $t$-dependence of the cross section in each energy bin separately. 
The detailed results are shown in Figs.~\ref{fig:overall} and~\ref{figs:EicC_dsigmadt_combined_plots} in Appendix~\ref{sec:appendix}.

\begin{figure}[htbp]
    \centering
    \includegraphics[width=0.92\linewidth]{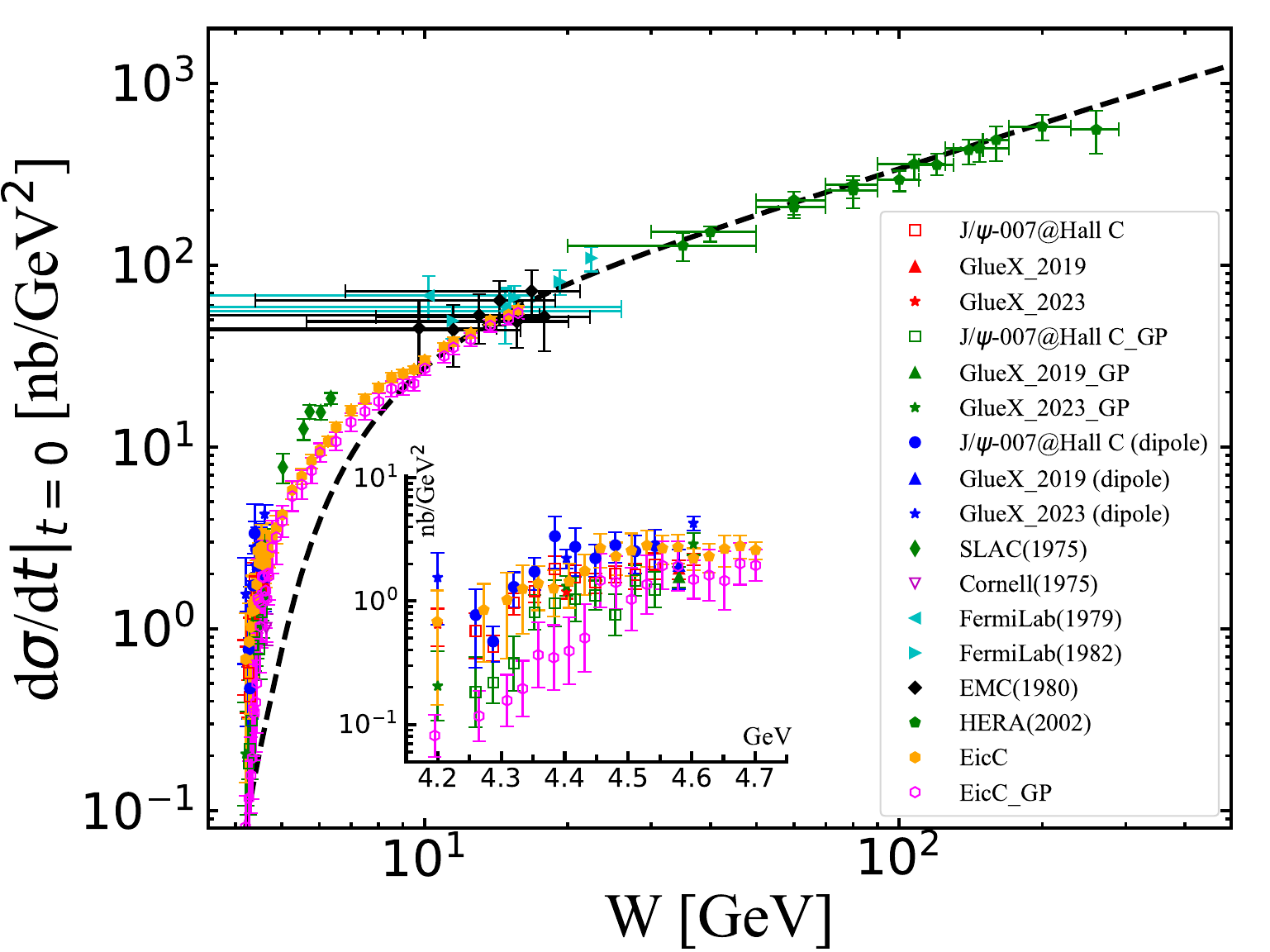}
    \caption{The comparison of the differential cross section $d\sigma/dt (t=0)$ of $\gamma p \to J/\psi p$.
    The orange and magenta data points are obtained within exponential function and one-dimensional GP from the pseudodata at EicC, respectively . 
    The blue data points are extrapolation of GlueX data by dipole formula, and the other experimental data are from \cite{Duran:2022xag,GlueX:2019mkq,GlueX:2023pev,Camerini:1975cy,Gittelman:1975ix,EuropeanMuon:1979nky,ZEUS:2002wfj,Binkley:1981kv} extrapolating by exponential function.
    The black dashed line is the contribution of imaginary part of amplitude through a once-subtracted dispersion relation \cite{Gryniuk:2016mpk}. 
    The inset is an enlarge of near threshold region. }
    \label{fig:W_dsigmadt_0}
\end{figure}

For comparison, in Fig.~\ref{fig:W_dsigmadt_0},
we also fit the entire data with the exponential $t$-dependence $d\sigma/dt \propto e^{-bt}$, 
and a dipole form \cite{Kharzeev:2021qkd,Duran:2022xag}, $d\sigma/dt \propto (1-t/\Lambda^2)^{-2}$,
specifically for JLab data.

It shall be noted that the systematic errors of extrapolation at high energies are negligible since the accessible minimum $t$ value is small,
as can be seen in Eq.~\eqref{eq:trange}.
For instance, $- t_0 < 0.01~\mathrm{GeV}^2$ above $W = 10$ GeV in comparison to $- t_0 = M_V^2M_N/(M_V+M_N) \simeq 2.23~\mathrm{GeV}^2$ at threshold.
The near-threshold errors quantified in GP is bigger than those in parametric method due to more flexible modelling extrapolation behaviour in GP without making strong assumptions about the functional form.
This larger uncertainty is not a drawback, 
but rather a more realistic quantification of our ignorance about the functional form of the cross-section in the unmeasured region. 
Parametric methods, by assuming a rigid function (e.g., exponential), 
can artificially underestimate the true extrapolation error.

The central values in GP are generally smaller than those in parametric method as shown in the enlarged inset of near threshold region in Fig. \ref{fig:W_dsigmadt_0}.
In any case, the near-threshold data cannot be explained alone by the contribution of imaginary part of amplitude through a once-subtracted dispersion relation \cite{Gryniuk:2016mpk} as shown by the black dashed line in Fig. \ref{fig:W_dsigmadt_0}, 
even after considering the uncertainties of extrapolation. 
This gap between data and imaginary part of amplitude indicates a clear evidence of real part of scattering amplitude.
This conclusion is robust, as it holds even when considering the more conservative, 
larger uncertainties provided by our model-independent GP extrapolation. 
This strengthens the evidence for a significant real part of the scattering amplitude near the threshold.

\begin{figure}
  \centering
    \includegraphics[width=0.92\linewidth]{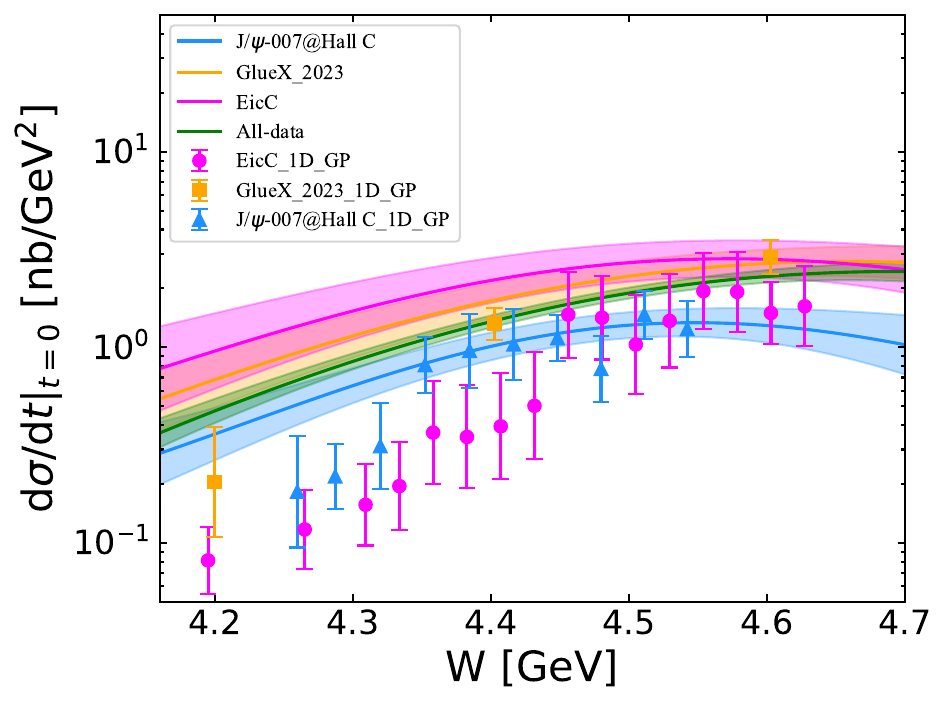}
  \caption{The differential cross section $d\sigma/dt (t=0)$ as a function of the $W$ with the shaded areas indicating the $1\sigma$ confidence band obtained within the two-dimensional GP. The results of one-dimensional GP for each dataset are also shown for comparison.}
  \label{figs:GP_dsigma_dt_Wt}
\end{figure}

We then perform a two-dimensional GP regression, 
using both $W$ and $t$ as joint input variables to analyze all datasets simultaneously. 
The optimized hyperparameters are listed in Table~\ref{tab:parameters}. 
The resulting prediction for the forward differential cross-section within the kinematic range $W \in [4.16,4.62]$ GeV and $|t| \in [0.46,2.78]~\mathrm{GeV}^2$ is shown in Fig.~\ref{figs:GP_dsigma_dt_Wt}.

\begin{table}[b]
\centering
\caption{The hyperparameters for different datasets of exclusive $J/\psi$ photoproduction. 
$l_1, l_2$ represent the correlation length-scales in the $W$ and $t$, respectively, 
while $\sigma^2$ is the signal variance.
The pseudo-data of EicC below W = 4.62 GeV are included in the analysis.}
\begin{tabular}{lcccccc}
\hline
         & $J/\psi$-007 & GlueX(2023)  & EicC & All data \\
\hline
$\sigma$ & 1.4439 & 2.5881  & 2.0226 & 4.2793  \\
$l_1$ ($W$) & 0.3784 & 0.7107  & 0.5618  & 0.8440 \\
$l_2$ ($t$) & 2.5538 & 3.2144  & 2.6974 &  7.4543 \\
\hline
\end{tabular}
\label{tab:parameters}
\end{table}

The results derived from separate datasets are consistent with each other, 
indicating the consistency of different experiments and the robustness of GP extrapolation. 
However, they exhibit significant uncertainties as indicated by the $1\sigma$ confidence band.
The central values in two-dimensional GP are larger than those in one-dimensional case, 
thus more consistent with parametric method.
As expected, a combined analysis of all datasets reduces the final uncertainty, 
underscoring the importance of future high-precision measurements.

Extracting physical quantities from the forward cross section introduces model dependence inevitably. 
Under the Vector Meson Dominance (VMD) assumption, 
the forward cross section $\gamma p \to J/\psi p$ is related to the forward $J/\psi-p$ elastic scattering amplitude $|T_{\psi p}|$ via \cite{Gryniuk:2016mpk}:
\be \label{eq:scattering}
\left. \frac{d\sigma}{dt} \right|_{t=0} = \left( \frac{e f_{V}}{M_{V}} \right)^2 \frac{1}{16 \pi \lambda^2(W^2,M_V^2,M_N^2)} \left| T_{\psi p} \right|^2 \,
\ee
with $e$ and $f_{V} = 0.278$ GeV being electric charge and $J/\psi$ decay constant, respectively.

\begin{figure}
    \centering
    \includegraphics[width=0.92\linewidth]{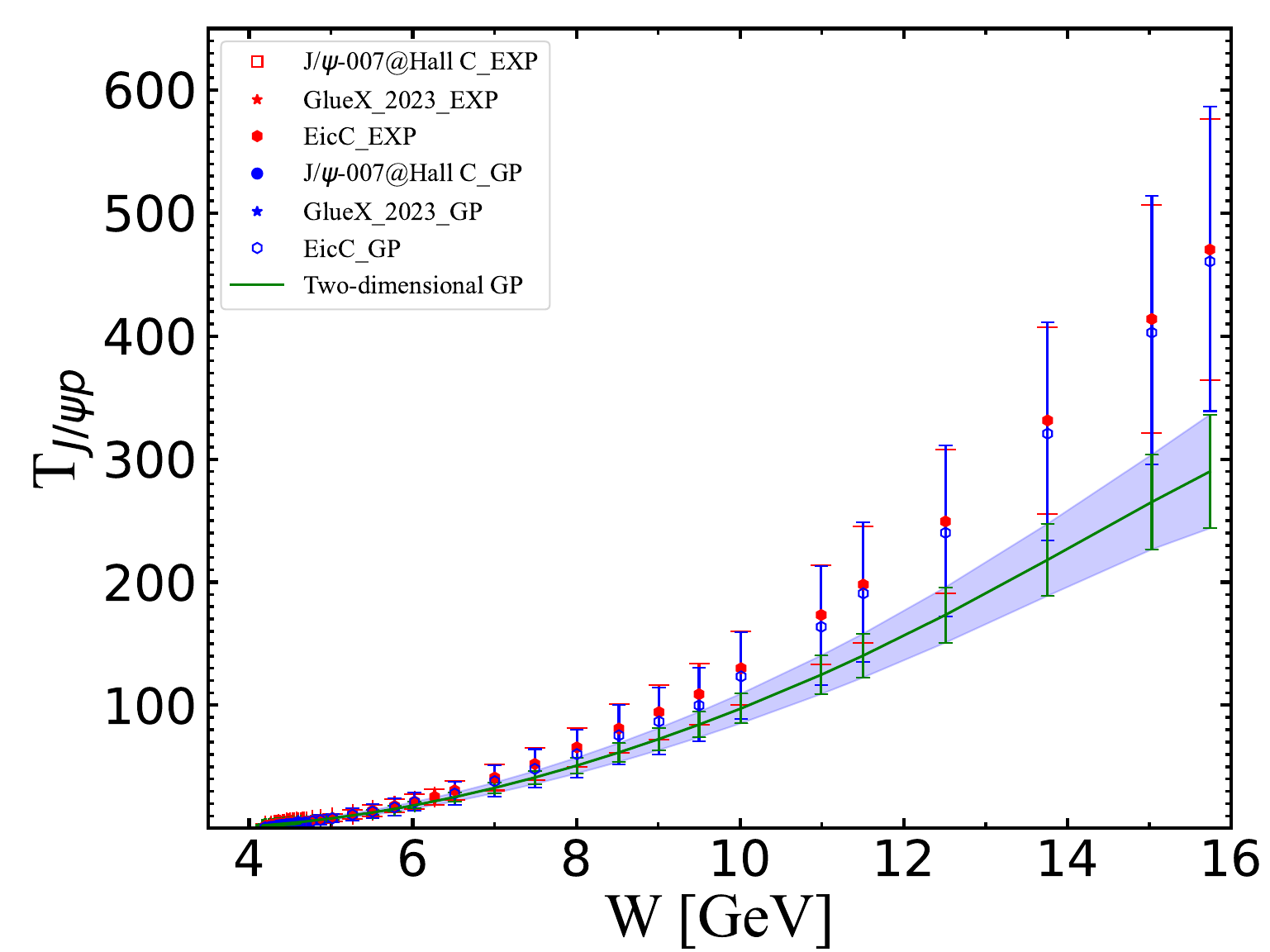}
    \caption{The absolute value of forward amplitude of the elastic $J/\psi p$ scattering as a function of the center-of-mass energy $W$ under the VMD assumption. The red points are extrapolation with exponential function and the blue points are those from GP.}
\label{fig:T_Comparison_exp_GP_twodimGP}
\end{figure}

Fig.~\ref{fig:T_Comparison_exp_GP_twodimGP} shows the extracted $|T_{\psi p}|$ with uncertainties by extrapolation of GP and exponential function. 
As expected, the uncertainties from the GP method (blue points) are larger than those from the exponential extrapolation (red points), 
since the latter assumes a exponential form a priori. 
While these near-threshold amplitudes could in principle be used to extract the $s$-wave $J/\psi-p$ scattering length by dispersion
relation\cite{Gryniuk:2016mpk}, 
we do not pursue this direction for two main reasons. 
First, the current data precision is insufficient for a reliable extraction. 
Second, the procedure itself carries significant theoretical uncertainties, 
including the validity of the VMD model at the large $|t_0|$ values encountered near threshold \cite{Du:2020bqj,Kharzeev:2021qkd},
and the uncertainties arising from dispersion relations.

The cross-section of near-threshold exclusive heavy quarkonium productions can be factorized in term of a form factor $G(t,\xi)$ 
and the non-relativistic wave function $\psi_{NR}(0)$ of heavy quarkonium at origin \cite{Guo:2021ibg,Guo:2023pqw,Guo:2023qgu}:
\be \label{eq:GFFcalc}
\frac{d\sigma}{dt} = \frac{\alpha_{em} e_Q^2}{4(W^2 - M_N^2)^2} \frac{(16\pi\alpha_s)^2}{3M_V^3} |\psi_{NR}(0)|^2 |G(t,\xi)|^2
\ee
in the domain of large-$t$ and big skewness with a definition of
\be
\xi = \frac{t - M_{V}^2}{2 M_N^2 + M_{V}^2 -t - 2 W^2}
\ee
The form factor $G(t,\xi)$ carries the information of the gluon GPDs of nucleon and $|G(t=0,\xi)|$ can be directly extracted from the data in Fig. \ref{fig:W_dsigmadt_0}.
It is pointed out that the scale anomaly of QCD enables the extraction of the proton mass radius from the form factor of the energy-momentum tensor trace \cite{Kharzeev:2021qkd}.
\footnote{
Note that physical explanation of mass radius is in fact of little relevant to the purpose here, 
considering that our main concern is the error of derivative of the slope of $G(t = 0)$, 
related directly to the observable $d\sigma/dt$ by Eq.\eqref{eq:GFFcalc}.
}

\begin{figure}
    \centering
    \includegraphics[width=0.92\linewidth]{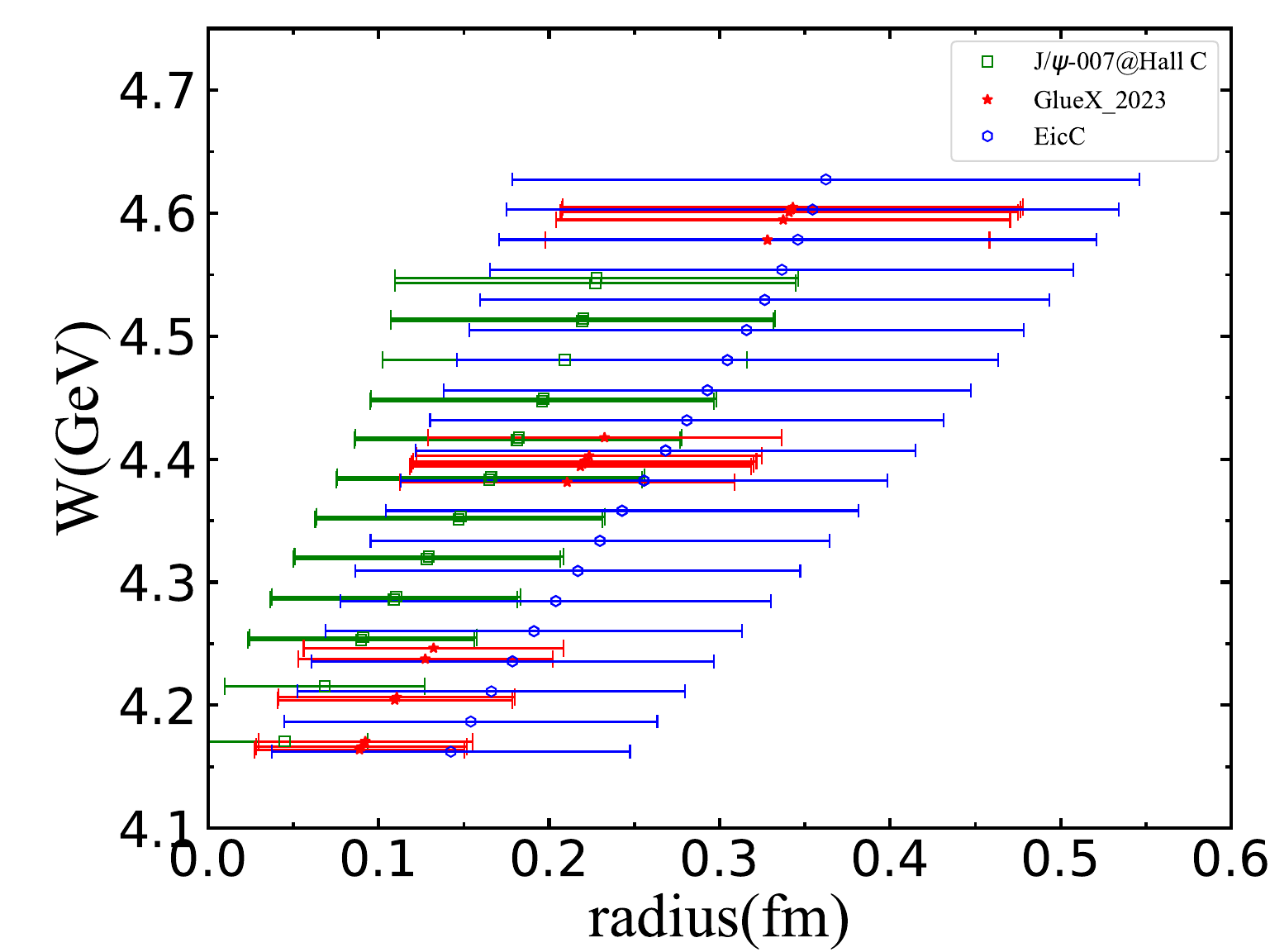}
    \caption{The proton mass radius extracted from near-threshold data at different center-of-mass energies $W$. 
    The horizontal bars represent the $1\sigma$ uncertainty derived from the GP extrapolations.}
    \label{fig:R_W}
\end{figure}

Specifically, the root-mean-square (RMS) radius of proton mass is the derivative of the scalar gravitational form factor at zero momentum transfer 
\be
\langle R_m^2 \rangle = \left. \frac{6}{M_N} \frac{dG(t)}{dt} \right|_{t=0},
\ee
where $G(t)$ is the scalar gravitational form factor that reflects the distribution of mass within the proton.
In order to satisfy the requirement of QCD multipole expansion, 
the data of threshold photoproduction of heavy quarkonia in the range of $W < 4.62$ GeV are used. 
The extracted values of mass radius scatter over a wide range, 
though they are still consistent with each other within large errors as shown in Fig.~\ref{fig:R_W}.
Our results can be contrasted with the values 0.55$\pm$0.03 fm \cite{Kharzeev:2021qkd} and 0.77$\pm$0.07 fm \cite{Guo:2023pqw} under the assumption of dipole $t$-dependence form factors, 
and 0.755$\pm$0.035 fm and 0.472$\pm$0.042 fm with tripole form factors in holographic and GPD approaches, respectively.
The calculation of the gravitational form factors through matrix elements of energy-momentum tensor within lattice QCD and dispersion relations gives 0.7464$\pm$0.025 fm \cite{Hackett:2023rif} and 0.70$^{+0.03}_{-0.04}$ fm \cite{Cao:2024zlf,Cao:2025dkv}, respectively.

The wide scatter of the results across different energy bins demonstrates that the proton mass radius cannot be conclusively determined with the current data precision, 
if the systematic uncertainty from the extrapolation is properly taken into account.
This inconclusiveness is a direct consequence of the large, 
systematically-driven uncertainty in the derivative extrapolation, 
which is honestly quantified by the GP and visualized in the error bars of Fig.~\ref{fig:R_W}.
The radius extracted from the pseudo-data at EicC is consistent with that based on data at JLab within errors.
This can be seemed as an validation of the reliability of the our procedure.
Bayesian inference with GP, instead of multi-pole $t$-dependence, combing with lattice QCD data is interesting in the future work \cite{Guo:2025jiz}.

\section{Summary} \label{sec:sum}

Probing the physics of exclusive charmonium photoproduction requires a robust extrapolation of the differential cross section to the forward regime, a region that is experimentally challenging to access. 
Traditional analyses, which rely on rigid parametric functions like exponential or multi-pole forms, 
often overlook the systematic uncertainties inherent in such extrapolations.

In this work, we addressed this challenge by employing Gaussian Process (GP) regression, 
a powerful non-parametric framework that provides a principled quantification of uncertainties. 
By constructing a two-dimensional covariance kernel in both squared momentum transfer $t$ and center-of-mass energy $W$, 
we performed a model-independent extrapolation of the world's $J/\psi$ photoproduction data to the $t\to 0$ and $W \to $ threshold region.
Our GP approach provides a more realistic estimate of the extrapolation uncertainty compared to traditional parametric methods by avoiding rigid assumptions about the functional form of the cross section.
Furthermore, we validated our framework by showing that the proton mass radius extracted from EicC pseudo-data are fully consistent with those from the original JLab measurements, as shown in Fig.~\ref{fig:R_W}. 
This cross-verification underscores the reliability and robustness of our method.

Future measurements with higher precision and broader kinematic coverage will further enhance the robustness of GP-based extrapolation. Owing to its ability to adaptively integrate prior knowledge with observed data, 
GP exhibits superior adaptability to small-sample, high-noise scenarios compared to parametric methods, 
which will effectively evaluate systematic extrapolation errors. 
Current datasets of moderate precision, when analyzed via GP extrapolation, 
already reveal features of the real part of the $J/\psi p$ production mechanism near the threshold region. 
This enables stringent tests of theoretical models and their domain of validity, 
such as those calculating scattering lengths under the VMD assumption or extracting the mass radius from the gravitational form factor.

Future potential applications include incorporating prior physical constraints, 
such as dispersion relations, directly into the GP prior, 
and extending our analysis to other exclusive processes,
such as $\Upsilon$ production \cite{Cao:2019gqo} or three dimensional cases \cite{Dutrieux:2021nlz,Cao:2023wyz}.
Furthermore, systematic comparisons with other modern non-parametric techniques, 
such as neural networks \cite{Graczyk:2014lba}, 
the statistical Schlessinger Point Method \cite{Cui:2021vgm}, 
and $z$-expansions \cite{Ye:2017gyb}, would also be an interesting avenue for future research.

\bigskip

\begin{acknowledgments}

X. C. is supported by the National Key R\&D Program of China under Grant No. 2023YFA1606703 and
the National Natural Science Foundation of China  (Grants No. 12075289).
Y. L. is supported by the National Natural Science Foundation of China  (Grants No. 12405105).
W. X. is supported by the Shandong Province Natural Science Foundation under Grant No. 2023HWYQ-010 and the National Key Research and Development Project of China (Contract No. 2023YFA1606800).

\end{acknowledgments}

\appendix

\section{Appendix} \label{sec:appendix}

\begin{figure*}[htbp]
   \centering
  {\includegraphics[width=0.88\linewidth]{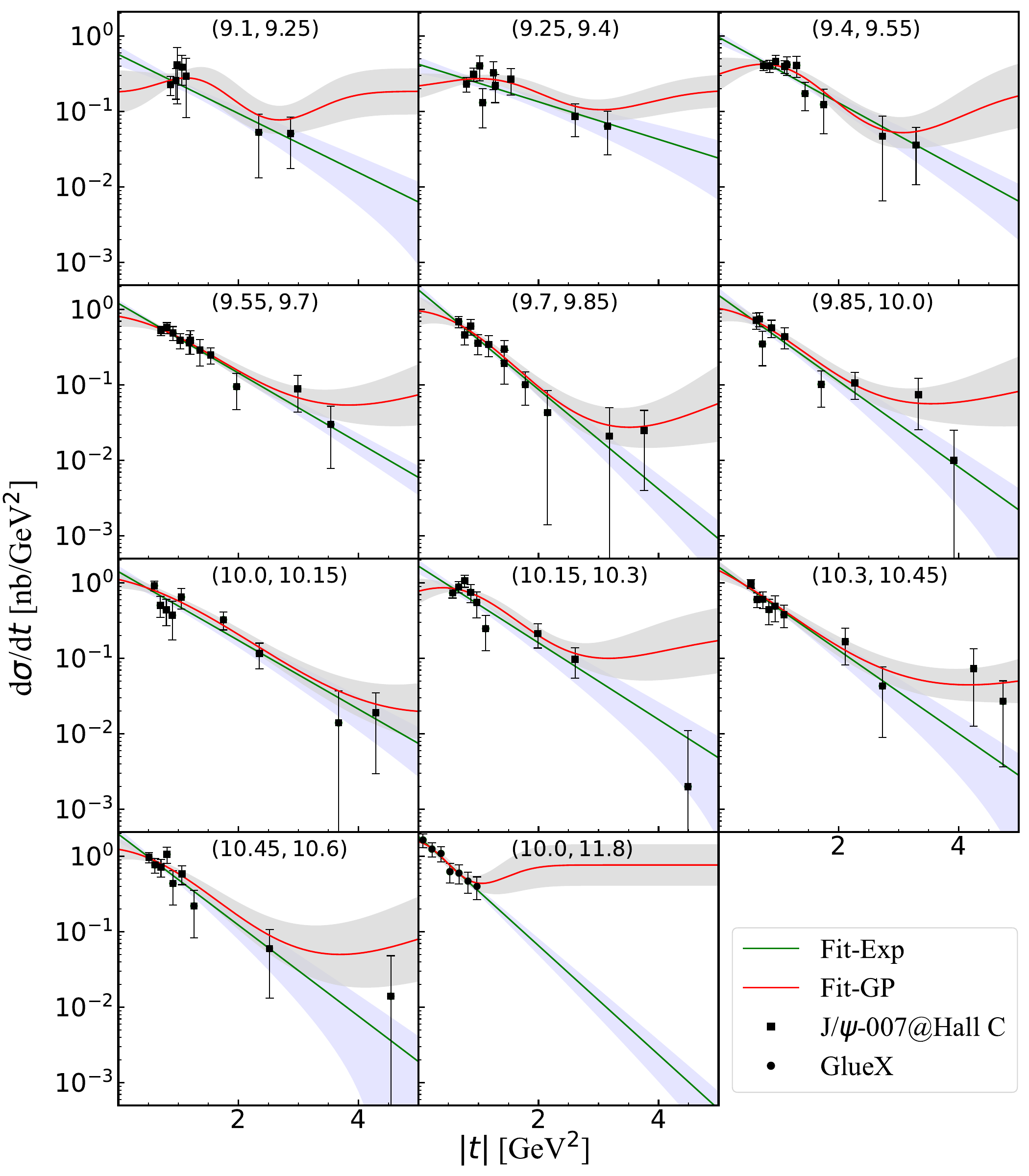}}
  {\includegraphics[width=0.83\linewidth]{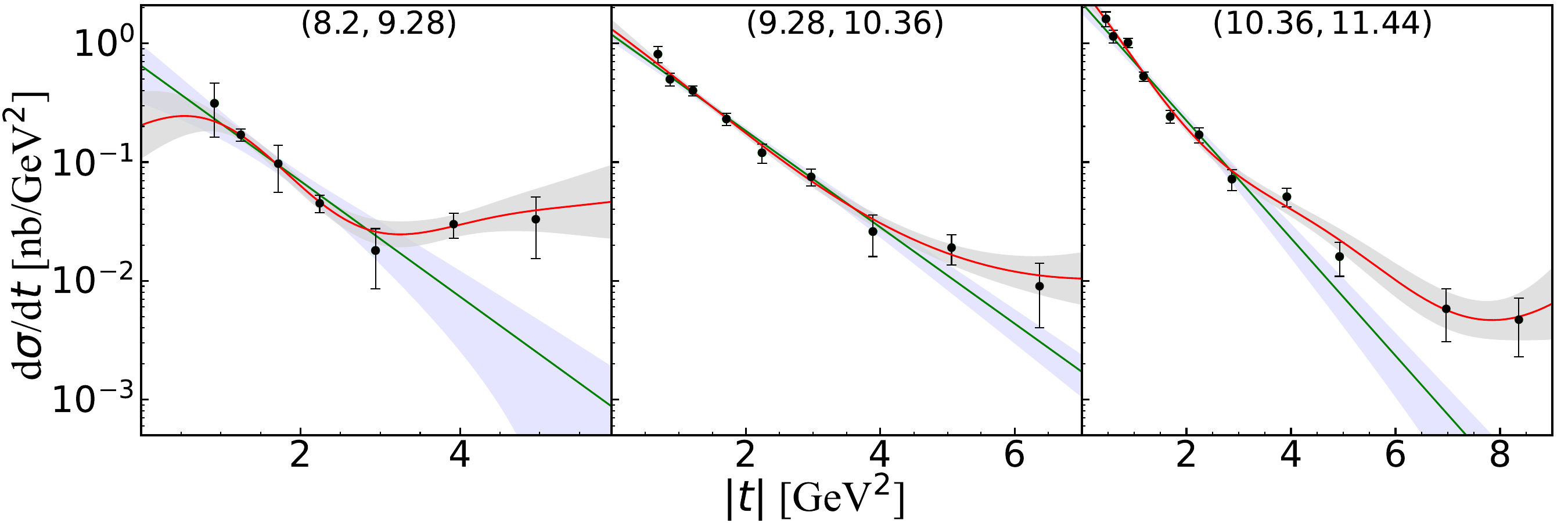}}
   \caption{The differential cross sections of the $J/\psi$ photo-production as a function of $-t$. The numerical values inside the parentheses in each subplots indicate the range of photon energy in the rest frame of the initial proton $E_{\gamma} = (W^2 - M_{p}^2)/{2 M_{p}}$ in the unit of GeV. 
   The experimental data are from GlueX (black circles) \cite{GlueX:2019mkq,GlueX:2023pev} and $J/\psi$-007 (black squares) \cite{Duran:2022xag} at JLab. 
   The solid red lines are fom the GP in the paper, and the solid green lines correspond to the results obtained from fitting with an exponential function $e^{-bt}$.}
   \label{fig:overall} 
\end{figure*}

\begin{figure*}[htbp]
   \centering
   {\includegraphics[width=0.92\linewidth]{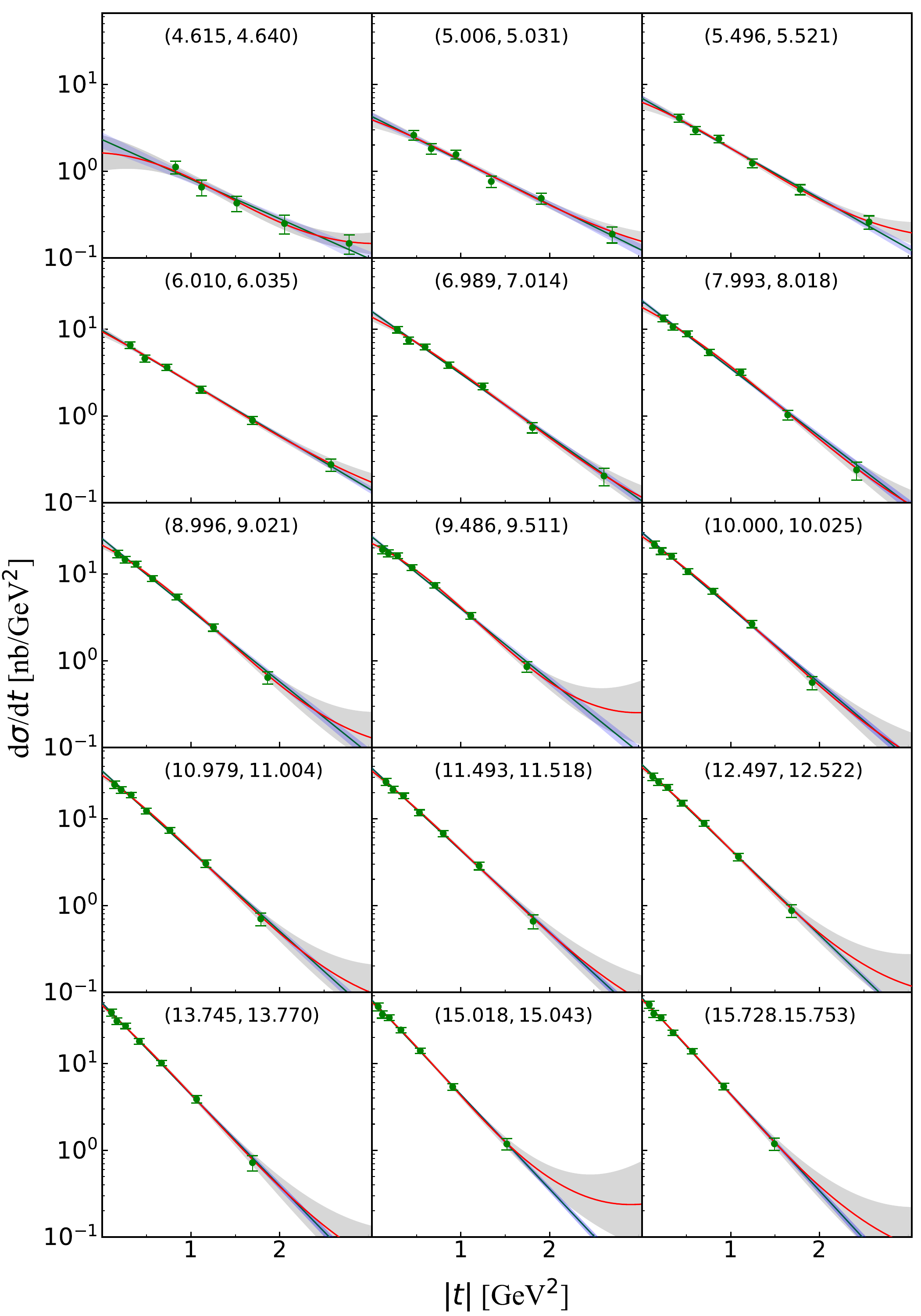}}
   \caption{The differential cross sections of the $J/\psi$ photo-production as a function of $-t$ for $Q^2 <$ 1.0 GeV$^2$ in selected $W$ bins as indicated inside the parentheses on the subplot. The green solid circles denote the pseudo-data of EicC. The meaning of solid lines are the same as those in Fig. \ref{fig:overall}.
}
   \label{figs:EicC_dsigmadt_combined_plots}
\end{figure*}

% \bigskip

\clearpage

\bibliography{HFatEicC_ref.bib}

\begin{thebibliography}{40}%
\makeatletter
\providecommand \@ifxundefined [1]{%
 \@ifx{#1\undefined}
}%
\providecommand \@ifnum [1]{%
 \ifnum #1\expandafter \@firstoftwo
 \else \expandafter \@secondoftwo
 \fi
}%
\providecommand \@ifx [1]{%
 \ifx #1\expandafter \@firstoftwo
 \else \expandafter \@secondoftwo
 \fi
}%
\providecommand \natexlab [1]{#1}%
\providecommand \enquote  [1]{``#1''}%
\providecommand \bibnamefont  [1]{#1}%
\providecommand \bibfnamefont [1]{#1}%
\providecommand \citenamefont [1]{#1}%
\providecommand \href@noop [0]{\@secondoftwo}%
\providecommand \href [0]{\begingroup \@sanitize@url \@href}%
\providecommand \@href[1]{\@@startlink{#1}\@@href}%
\providecommand \@@href[1]{\endgroup#1\@@endlink}%
\providecommand \@sanitize@url [0]{\catcode `\\12\catcode `\$12\catcode `\&12\catcode `\#12\catcode `\^12\catcode `\_12\catcode `\%12\relax}%
\providecommand \@@startlink[1]{}%
\providecommand \@@endlink[0]{}%
\providecommand \url  [0]{\begingroup\@sanitize@url \@url }%
\providecommand \@url [1]{\endgroup\@href {#1}{\urlprefix }}%
\providecommand \urlprefix  [0]{URL }%
\providecommand \Eprint [0]{\href }%
\providecommand \doibase [0]{http://dx.doi.org/}%
\providecommand \selectlanguage [0]{\@gobble}%
\providecommand \bibinfo  [0]{\@secondoftwo}%
\providecommand \bibfield  [0]{\@secondoftwo}%
\providecommand \translation [1]{[#1]}%
\providecommand \BibitemOpen [0]{}%
\providecommand \bibitemStop [0]{}%
\providecommand \bibitemNoStop [0]{.\EOS\space}%
\providecommand \EOS [0]{\spacefactor3000\relax}%
\providecommand \BibitemShut  [1]{\csname bibitem#1\endcsname}%
\let\auto@bib@innerbib\@empty
%</preamble>
\bibitem [{\citenamefont {Diehl}(2003)}]{Diehl:2003ny}%
  \BibitemOpen
  \bibfield  {author} {\bibinfo {author} {\bibfnamefont {M.}~\bibnamefont {Diehl}},\ }\href {\doibase 10.1016/j.physrep.2003.08.002} {\bibfield  {journal} {\bibinfo  {journal} {Phys. Rept.}\ }\textbf {\bibinfo {volume} {388}},\ \bibinfo {pages} {41} (\bibinfo {year} {2003})},\ \Eprint {http://arxiv.org/abs/hep-ph/0307382} {arXiv:hep-ph/0307382} \BibitemShut {NoStop}%
\bibitem [{\citenamefont {Belitsky}\ and\ \citenamefont {Radyushkin}(2005)}]{Belitsky:2005qn}%
  \BibitemOpen
  \bibfield  {author} {\bibinfo {author} {\bibfnamefont {A.~V.}\ \bibnamefont {Belitsky}}\ and\ \bibinfo {author} {\bibfnamefont {A.~V.}\ \bibnamefont {Radyushkin}},\ }\href {\doibase 10.1016/j.physrep.2005.06.002} {\bibfield  {journal} {\bibinfo  {journal} {Phys. Rept.}\ }\textbf {\bibinfo {volume} {418}},\ \bibinfo {pages} {1} (\bibinfo {year} {2005})},\ \Eprint {http://arxiv.org/abs/hep-ph/0504030} {arXiv:hep-ph/0504030} \BibitemShut {NoStop}%
\bibitem [{\citenamefont {Guo}\ \emph {et~al.}(2021)\citenamefont {Guo}, \citenamefont {Ji},\ and\ \citenamefont {Liu}}]{Guo:2021ibg}%
  \BibitemOpen
  \bibfield  {author} {\bibinfo {author} {\bibfnamefont {Y.}~\bibnamefont {Guo}}, \bibinfo {author} {\bibfnamefont {X.}~\bibnamefont {Ji}}, \ and\ \bibinfo {author} {\bibfnamefont {Y.}~\bibnamefont {Liu}},\ }\href {\doibase 10.1103/PhysRevD.103.096010} {\bibfield  {journal} {\bibinfo  {journal} {Phys. Rev. D}\ }\textbf {\bibinfo {volume} {103}},\ \bibinfo {pages} {096010} (\bibinfo {year} {2021})},\ \Eprint {http://arxiv.org/abs/2103.11506} {arXiv:2103.11506 [hep-ph]} \BibitemShut {NoStop}%
\bibitem [{\citenamefont {Gryniuk}\ and\ \citenamefont {Vanderhaeghen}(2016)}]{Gryniuk:2016mpk}%
  \BibitemOpen
  \bibfield  {author} {\bibinfo {author} {\bibfnamefont {O.}~\bibnamefont {Gryniuk}}\ and\ \bibinfo {author} {\bibfnamefont {M.}~\bibnamefont {Vanderhaeghen}},\ }\href {\doibase 10.1103/PhysRevD.94.074001} {\bibfield  {journal} {\bibinfo  {journal} {Phys. Rev. D}\ }\textbf {\bibinfo {volume} {94}},\ \bibinfo {pages} {074001} (\bibinfo {year} {2016})},\ \Eprint {http://arxiv.org/abs/1608.08205} {arXiv:1608.08205 [hep-ph]} \BibitemShut {NoStop}%
\bibitem [{\citenamefont {Cao}\ and\ \citenamefont {Dai}(2019)}]{Cao:2019kst}%
  \BibitemOpen
  \bibfield  {author} {\bibinfo {author} {\bibfnamefont {X.}~\bibnamefont {Cao}}\ and\ \bibinfo {author} {\bibfnamefont {J.-p.}\ \bibnamefont {Dai}},\ }\href {\doibase 10.1103/PhysRevD.100.054033} {\bibfield  {journal} {\bibinfo  {journal} {Phys. Rev. D}\ }\textbf {\bibinfo {volume} {100}},\ \bibinfo {pages} {054033} (\bibinfo {year} {2019})},\ \Eprint {http://arxiv.org/abs/1904.06015} {arXiv:1904.06015 [hep-ph]} \BibitemShut {NoStop}%
\bibitem [{\citenamefont {Winney}\ \emph {et~al.}(2023)\citenamefont {Winney} \emph {et~al.}}]{JointPhysicsAnalysisCenter:2023qgg}%
  \BibitemOpen
  \bibfield  {author} {\bibinfo {author} {\bibfnamefont {D.}~\bibnamefont {Winney}} \emph {et~al.} (\bibinfo {collaboration} {Joint Physics Analysis Center}),\ }\href {\doibase 10.1103/PhysRevD.108.054018} {\bibfield  {journal} {\bibinfo  {journal} {Phys. Rev. D}\ }\textbf {\bibinfo {volume} {108}},\ \bibinfo {pages} {054018} (\bibinfo {year} {2023})},\ \Eprint {http://arxiv.org/abs/2305.01449} {arXiv:2305.01449 [hep-ph]} \BibitemShut {NoStop}%
\bibitem [{\citenamefont {Guo}\ \emph {et~al.}(2024)\citenamefont {Guo}, \citenamefont {Ji},\ and\ \citenamefont {Yuan}}]{Guo:2023qgu}%
  \BibitemOpen
  \bibfield  {author} {\bibinfo {author} {\bibfnamefont {Y.}~\bibnamefont {Guo}}, \bibinfo {author} {\bibfnamefont {X.}~\bibnamefont {Ji}}, \ and\ \bibinfo {author} {\bibfnamefont {F.}~\bibnamefont {Yuan}},\ }\href {\doibase 10.1103/PhysRevD.109.014014} {\bibfield  {journal} {\bibinfo  {journal} {Phys. Rev. D}\ }\textbf {\bibinfo {volume} {109}},\ \bibinfo {pages} {014014} (\bibinfo {year} {2024})},\ \Eprint {http://arxiv.org/abs/2308.13006} {arXiv:2308.13006 [hep-ph]} \BibitemShut {NoStop}%
\bibitem [{\citenamefont {Sun}\ \emph {et~al.}(2022)\citenamefont {Sun}, \citenamefont {Tong},\ and\ \citenamefont {Yuan}}]{Sun:2021pyw}%
  \BibitemOpen
  \bibfield  {author} {\bibinfo {author} {\bibfnamefont {P.}~\bibnamefont {Sun}}, \bibinfo {author} {\bibfnamefont {X.-B.}\ \bibnamefont {Tong}}, \ and\ \bibinfo {author} {\bibfnamefont {F.}~\bibnamefont {Yuan}},\ }\href {\doibase 10.1103/PhysRevD.105.054032} {\bibfield  {journal} {\bibinfo  {journal} {Phys. Rev. D}\ }\textbf {\bibinfo {volume} {105}},\ \bibinfo {pages} {054032} (\bibinfo {year} {2022})},\ \Eprint {http://arxiv.org/abs/2111.07034} {arXiv:2111.07034 [hep-ph]} \BibitemShut {NoStop}%
\bibitem [{\citenamefont {Sun}\ \emph {et~al.}(2021)\citenamefont {Sun}, \citenamefont {Tong},\ and\ \citenamefont {Yuan}}]{Sun:2021gmi}%
  \BibitemOpen
  \bibfield  {author} {\bibinfo {author} {\bibfnamefont {P.}~\bibnamefont {Sun}}, \bibinfo {author} {\bibfnamefont {X.-B.}\ \bibnamefont {Tong}}, \ and\ \bibinfo {author} {\bibfnamefont {F.}~\bibnamefont {Yuan}},\ }\href {\doibase 10.1016/j.physletb.2021.136655} {\bibfield  {journal} {\bibinfo  {journal} {Phys. Lett. B}\ }\textbf {\bibinfo {volume} {822}},\ \bibinfo {pages} {136655} (\bibinfo {year} {2021})},\ \Eprint {http://arxiv.org/abs/2103.12047} {arXiv:2103.12047 [hep-ph]} \BibitemShut {NoStop}%
\bibitem [{\citenamefont {Kharzeev}(2021)}]{Kharzeev:2021qkd}%
  \BibitemOpen
  \bibfield  {author} {\bibinfo {author} {\bibfnamefont {D.~E.}\ \bibnamefont {Kharzeev}},\ }\href {\doibase 10.1103/PhysRevD.104.054015} {\bibfield  {journal} {\bibinfo  {journal} {Phys. Rev. D}\ }\textbf {\bibinfo {volume} {104}},\ \bibinfo {pages} {054015} (\bibinfo {year} {2021})},\ \Eprint {http://arxiv.org/abs/2102.00110} {arXiv:2102.00110 [hep-ph]} \BibitemShut {NoStop}%
\bibitem [{\citenamefont {Guo}\ \emph {et~al.}(2025)\citenamefont {Guo}, \citenamefont {Yuan},\ and\ \citenamefont {Zhao}}]{Guo:2025jiz}%
  \BibitemOpen
  \bibfield  {author} {\bibinfo {author} {\bibfnamefont {Y.}~\bibnamefont {Guo}}, \bibinfo {author} {\bibfnamefont {F.}~\bibnamefont {Yuan}}, \ and\ \bibinfo {author} {\bibfnamefont {W.}~\bibnamefont {Zhao}},\ }\href {\doibase 10.1103/3x7r-ythq} {\bibfield  {journal} {\bibinfo  {journal} {Phys. Rev. Lett.}\ }\textbf {\bibinfo {volume} {135}},\ \bibinfo {pages} {111902} (\bibinfo {year} {2025})},\ \Eprint {http://arxiv.org/abs/2501.10532} {arXiv:2501.10532 [hep-ph]} \BibitemShut {NoStop}%
\bibitem [{\citenamefont {Rasmussen}\ and\ \citenamefont {Williams}(2006)}]{rasmussen:2006gau}%
  \BibitemOpen
  \bibfield  {author} {\bibinfo {author} {\bibfnamefont {C.~E.}\ \bibnamefont {Rasmussen}}\ and\ \bibinfo {author} {\bibfnamefont {C.~K.~I.}\ \bibnamefont {Williams}},\ }\href@noop {} {\emph {\bibinfo {title} {Gaussian Processes for Machine Learning}}},\ Adaptive Computation and Machine Learning\ (\bibinfo  {publisher} {MIT Press},\ \bibinfo {address} {Cambridge, Mass},\ \bibinfo {year} {2006})\BibitemShut {NoStop}%
\bibitem [{\citenamefont {Ferguson}\ \emph {et~al.}(2025)\citenamefont {Ferguson}, \citenamefont {Ireland},\ and\ \citenamefont {McKinnon}}]{Ferguson:2025ddu}%
  \BibitemOpen
  \bibfield  {author} {\bibinfo {author} {\bibfnamefont {R.~F.}\ \bibnamefont {Ferguson}}, \bibinfo {author} {\bibfnamefont {D.~G.}\ \bibnamefont {Ireland}}, \ and\ \bibinfo {author} {\bibfnamefont {B.}~\bibnamefont {McKinnon}},\ }\href@noop {} {\  (\bibinfo {year} {2025})},\ \Eprint {http://arxiv.org/abs/2505.01473} {arXiv:2505.01473 [physics.data-an]} \BibitemShut {NoStop}%
\bibitem [{\citenamefont {Jaiswal}\ \emph {et~al.}(2025)\citenamefont {Jaiswal}, \citenamefont {Shen}, \citenamefont {Furnstahl}, \citenamefont {Heinz},\ and\ \citenamefont {Pratola}}]{Jaiswal:2025hyp}%
  \BibitemOpen
  \bibfield  {author} {\bibinfo {author} {\bibfnamefont {S.}~\bibnamefont {Jaiswal}}, \bibinfo {author} {\bibfnamefont {C.}~\bibnamefont {Shen}}, \bibinfo {author} {\bibfnamefont {R.~J.}\ \bibnamefont {Furnstahl}}, \bibinfo {author} {\bibfnamefont {U.}~\bibnamefont {Heinz}}, \ and\ \bibinfo {author} {\bibfnamefont {M.~T.}\ \bibnamefont {Pratola}},\ }\href@noop {} {\  (\bibinfo {year} {2025})},\ \Eprint {http://arxiv.org/abs/2504.13144} {arXiv:2504.13144 [hep-ph]} \BibitemShut {NoStop}%
\bibitem [{\citenamefont {Fowlie}\ and\ \citenamefont {Li}(2023)}]{Fowlie:2023cta}%
  \BibitemOpen
  \bibfield  {author} {\bibinfo {author} {\bibfnamefont {A.}~\bibnamefont {Fowlie}}\ and\ \bibinfo {author} {\bibfnamefont {Q.}~\bibnamefont {Li}},\ }\href {\doibase 10.1140/epjc/s10052-023-12110-9} {\bibfield  {journal} {\bibinfo  {journal} {Eur. Phys. J. C}\ }\textbf {\bibinfo {volume} {83}},\ \bibinfo {pages} {943} (\bibinfo {year} {2023})},\ \Eprint {http://arxiv.org/abs/2306.17385} {arXiv:2306.17385 [hep-ph]} \BibitemShut {NoStop}%
\bibitem [{\citenamefont {Candido}\ \emph {et~al.}(2024)\citenamefont {Candido}, \citenamefont {Del~Debbio}, \citenamefont {Giani},\ and\ \citenamefont {Petrillo}}]{Candido:2024hjt}%
  \BibitemOpen
  \bibfield  {author} {\bibinfo {author} {\bibfnamefont {A.}~\bibnamefont {Candido}}, \bibinfo {author} {\bibfnamefont {L.}~\bibnamefont {Del~Debbio}}, \bibinfo {author} {\bibfnamefont {T.}~\bibnamefont {Giani}}, \ and\ \bibinfo {author} {\bibfnamefont {G.}~\bibnamefont {Petrillo}},\ }\href {\doibase 10.1140/epjc/s10052-024-13100-1} {\bibfield  {journal} {\bibinfo  {journal} {Eur. Phys. J. C}\ }\textbf {\bibinfo {volume} {84}},\ \bibinfo {pages} {716} (\bibinfo {year} {2024})},\ \Eprint {http://arxiv.org/abs/2404.07573} {arXiv:2404.07573 [hep-ph]} \BibitemShut {NoStop}%
\bibitem [{\citenamefont {Dutrieux}\ \emph {et~al.}(2025)\citenamefont {Dutrieux}, \citenamefont {Karpie}, \citenamefont {Monahan}, \citenamefont {Orginos}, \citenamefont {Radyushkin}, \citenamefont {Richards},\ and\ \citenamefont {Zafeiropoulos}}]{Dutrieux:2025jed}%
  \BibitemOpen
  \bibfield  {author} {\bibinfo {author} {\bibfnamefont {H.}~\bibnamefont {Dutrieux}}, \bibinfo {author} {\bibfnamefont {J.}~\bibnamefont {Karpie}}, \bibinfo {author} {\bibfnamefont {C.~J.}\ \bibnamefont {Monahan}}, \bibinfo {author} {\bibfnamefont {K.}~\bibnamefont {Orginos}}, \bibinfo {author} {\bibfnamefont {A.}~\bibnamefont {Radyushkin}}, \bibinfo {author} {\bibfnamefont {D.}~\bibnamefont {Richards}}, \ and\ \bibinfo {author} {\bibfnamefont {S.}~\bibnamefont {Zafeiropoulos}},\ }\href@noop {} {\  (\bibinfo {year} {2025})},\ \Eprint {http://arxiv.org/abs/2504.17706} {arXiv:2504.17706 [hep-lat]} \BibitemShut {NoStop}%
\bibitem [{\citenamefont {Chen}\ \emph {et~al.}(2025)\citenamefont {Chen} \emph {et~al.}}]{Chen:2025cxr}%
  \BibitemOpen
  \bibfield  {author} {\bibinfo {author} {\bibfnamefont {J.-W.}\ \bibnamefont {Chen}} \emph {et~al.},\ }\href@noop {} {\  (\bibinfo {year} {2025})},\ \Eprint {http://arxiv.org/abs/2505.14619} {arXiv:2505.14619 [hep-lat]} \BibitemShut {NoStop}%
\bibitem [{\citenamefont {Barr}\ and\ \citenamefont {Liu}(2025)}]{Barr:2025lba}%
  \BibitemOpen
  \bibfield  {author} {\bibinfo {author} {\bibfnamefont {J.}~\bibnamefont {Barr}}\ and\ \bibinfo {author} {\bibfnamefont {B.}~\bibnamefont {Liu}},\ }\href@noop {} {\  (\bibinfo {year} {2025})},\ \Eprint {http://arxiv.org/abs/2503.07289} {arXiv:2503.07289 [hep-ex]} \BibitemShut {NoStop}%
\bibitem [{\citenamefont {Cao}(2023)}]{Cao:2023rhu}%
  \BibitemOpen
  \bibfield  {author} {\bibinfo {author} {\bibfnamefont {X.}~\bibnamefont {Cao}},\ }\href {\doibase 10.1007/s11467-023-1264-8} {\bibfield  {journal} {\bibinfo  {journal} {Front. Phys. (Beijing)}\ }\textbf {\bibinfo {volume} {18}},\ \bibinfo {pages} {44600} (\bibinfo {year} {2023})},\ \Eprint {http://arxiv.org/abs/2301.11253} {arXiv:2301.11253 [hep-ph]} \BibitemShut {NoStop}%
\bibitem [{\citenamefont {Wang}\ \emph {et~al.}(2024)\citenamefont {Wang}, \citenamefont {Cao}, \citenamefont {Guo}, \citenamefont {Gong}, \citenamefont {Kang}, \citenamefont {Liang}, \citenamefont {Wu},\ and\ \citenamefont {Xie}}]{Wang:2023thy}%
  \BibitemOpen
  \bibfield  {author} {\bibinfo {author} {\bibfnamefont {X.}~\bibnamefont {Wang}}, \bibinfo {author} {\bibfnamefont {X.}~\bibnamefont {Cao}}, \bibinfo {author} {\bibfnamefont {A.}~\bibnamefont {Guo}}, \bibinfo {author} {\bibfnamefont {L.}~\bibnamefont {Gong}}, \bibinfo {author} {\bibfnamefont {X.-S.}\ \bibnamefont {Kang}}, \bibinfo {author} {\bibfnamefont {Y.-T.}\ \bibnamefont {Liang}}, \bibinfo {author} {\bibfnamefont {J.-J.}\ \bibnamefont {Wu}}, \ and\ \bibinfo {author} {\bibfnamefont {Y.-P.}\ \bibnamefont {Xie}},\ }\href {\doibase 10.1140/epjc/s10052-024-13033-9} {\bibfield  {journal} {\bibinfo  {journal} {Eur. Phys. J. C}\ }\textbf {\bibinfo {volume} {84}},\ \bibinfo {pages} {684} (\bibinfo {year} {2024})},\ \Eprint {http://arxiv.org/abs/2311.07008} {arXiv:2311.07008 [hep-ph]} \BibitemShut {NoStop}%
\bibitem [{\citenamefont {Ali}\ \emph {et~al.}(2019)\citenamefont {Ali} \emph {et~al.}}]{GlueX:2019mkq}%
  \BibitemOpen
  \bibfield  {author} {\bibinfo {author} {\bibfnamefont {A.}~\bibnamefont {Ali}} \emph {et~al.} (\bibinfo {collaboration} {GlueX}),\ }\href {\doibase 10.1103/PhysRevLett.123.072001} {\bibfield  {journal} {\bibinfo  {journal} {Phys. Rev. Lett.}\ }\textbf {\bibinfo {volume} {123}},\ \bibinfo {pages} {072001} (\bibinfo {year} {2019})},\ \Eprint {http://arxiv.org/abs/1905.10811} {arXiv:1905.10811 [nucl-ex]} \BibitemShut {NoStop}%
\bibitem [{\citenamefont {Adhikari}\ \emph {et~al.}(2023)\citenamefont {Adhikari} \emph {et~al.}}]{GlueX:2023pev}%
  \BibitemOpen
  \bibfield  {author} {\bibinfo {author} {\bibfnamefont {S.}~\bibnamefont {Adhikari}} \emph {et~al.} (\bibinfo {collaboration} {GlueX}),\ }\href {\doibase 10.1103/PhysRevC.108.025201} {\bibfield  {journal} {\bibinfo  {journal} {Phys. Rev. C}\ }\textbf {\bibinfo {volume} {108}},\ \bibinfo {pages} {025201} (\bibinfo {year} {2023})},\ \Eprint {http://arxiv.org/abs/2304.03845} {arXiv:2304.03845 [nucl-ex]} \BibitemShut {NoStop}%
\bibitem [{\citenamefont {Duran}\ \emph {et~al.}(2023)\citenamefont {Duran} \emph {et~al.}}]{Duran:2022xag}%
  \BibitemOpen
  \bibfield  {author} {\bibinfo {author} {\bibfnamefont {B.}~\bibnamefont {Duran}} \emph {et~al.},\ }\href {\doibase 10.1038/s41586-023-05730-4} {\bibfield  {journal} {\bibinfo  {journal} {Nature}\ }\textbf {\bibinfo {volume} {615}},\ \bibinfo {pages} {813} (\bibinfo {year} {2023})},\ \Eprint {http://arxiv.org/abs/2207.05212} {arXiv:2207.05212 [nucl-ex]} \BibitemShut {NoStop}%
\bibitem [{\citenamefont {Camerini}\ \emph {et~al.}(1975)\citenamefont {Camerini}, \citenamefont {Learned}, \citenamefont {Prepost}, \citenamefont {Spencer}, \citenamefont {Wiser}, \citenamefont {Ash}, \citenamefont {Anderson}, \citenamefont {Ritson}, \citenamefont {Sherden},\ and\ \citenamefont {Sinclair}}]{Camerini:1975cy}%
  \BibitemOpen
  \bibfield  {author} {\bibinfo {author} {\bibfnamefont {U.}~\bibnamefont {Camerini}}, \bibinfo {author} {\bibfnamefont {J.~G.}\ \bibnamefont {Learned}}, \bibinfo {author} {\bibfnamefont {R.}~\bibnamefont {Prepost}}, \bibinfo {author} {\bibfnamefont {C.~M.}\ \bibnamefont {Spencer}}, \bibinfo {author} {\bibfnamefont {D.~E.}\ \bibnamefont {Wiser}}, \bibinfo {author} {\bibfnamefont {W.}~\bibnamefont {Ash}}, \bibinfo {author} {\bibfnamefont {R.~L.}\ \bibnamefont {Anderson}}, \bibinfo {author} {\bibfnamefont {D.}~\bibnamefont {Ritson}}, \bibinfo {author} {\bibfnamefont {D.}~\bibnamefont {Sherden}}, \ and\ \bibinfo {author} {\bibfnamefont {C.~K.}\ \bibnamefont {Sinclair}},\ }\href {\doibase 10.1103/PhysRevLett.35.483} {\bibfield  {journal} {\bibinfo  {journal} {Phys. Rev. Lett.}\ }\textbf {\bibinfo {volume} {35}},\ \bibinfo {pages} {483} (\bibinfo {year} {1975})}\BibitemShut {NoStop}%
\bibitem [{\citenamefont {Gittelman}\ \emph {et~al.}(1975)\citenamefont {Gittelman}, \citenamefont {Hanson}, \citenamefont {Larson}, \citenamefont {Loh}, \citenamefont {Silverman},\ and\ \citenamefont {Theodosiou}}]{Gittelman:1975ix}%
  \BibitemOpen
  \bibfield  {author} {\bibinfo {author} {\bibfnamefont {B.}~\bibnamefont {Gittelman}}, \bibinfo {author} {\bibfnamefont {K.~M.}\ \bibnamefont {Hanson}}, \bibinfo {author} {\bibfnamefont {D.}~\bibnamefont {Larson}}, \bibinfo {author} {\bibfnamefont {E.}~\bibnamefont {Loh}}, \bibinfo {author} {\bibfnamefont {A.}~\bibnamefont {Silverman}}, \ and\ \bibinfo {author} {\bibfnamefont {G.}~\bibnamefont {Theodosiou}},\ }\href {\doibase 10.1103/PhysRevLett.35.1616} {\bibfield  {journal} {\bibinfo  {journal} {Phys. Rev. Lett.}\ }\textbf {\bibinfo {volume} {35}},\ \bibinfo {pages} {1616} (\bibinfo {year} {1975})}\BibitemShut {NoStop}%
\bibitem [{\citenamefont {Aubert}\ \emph {et~al.}(1980)\citenamefont {Aubert} \emph {et~al.}}]{EuropeanMuon:1979nky}%
  \BibitemOpen
  \bibfield  {author} {\bibinfo {author} {\bibfnamefont {J.~J.}\ \bibnamefont {Aubert}} \emph {et~al.} (\bibinfo {collaboration} {European Muon}),\ }\href {\doibase 10.1016/0370-2693(80)90027-1} {\bibfield  {journal} {\bibinfo  {journal} {Phys. Lett. B}\ }\textbf {\bibinfo {volume} {89}},\ \bibinfo {pages} {267} (\bibinfo {year} {1980})}\BibitemShut {NoStop}%
\bibitem [{\citenamefont {Chekanov}\ \emph {et~al.}(2002)\citenamefont {Chekanov} \emph {et~al.}}]{ZEUS:2002wfj}%
  \BibitemOpen
  \bibfield  {author} {\bibinfo {author} {\bibfnamefont {S.}~\bibnamefont {Chekanov}} \emph {et~al.} (\bibinfo {collaboration} {ZEUS}),\ }\href {\doibase 10.1007/s10052-002-0953-7} {\bibfield  {journal} {\bibinfo  {journal} {Eur. Phys. J. C}\ }\textbf {\bibinfo {volume} {24}},\ \bibinfo {pages} {345} (\bibinfo {year} {2002})},\ \Eprint {http://arxiv.org/abs/hep-ex/0201043} {arXiv:hep-ex/0201043} \BibitemShut {NoStop}%
\bibitem [{\citenamefont {Binkley}\ \emph {et~al.}(1982)\citenamefont {Binkley} \emph {et~al.}}]{Binkley:1981kv}%
  \BibitemOpen
  \bibfield  {author} {\bibinfo {author} {\bibfnamefont {M.~E.}\ \bibnamefont {Binkley}} \emph {et~al.},\ }\href {\doibase 10.1103/PhysRevLett.48.73} {\bibfield  {journal} {\bibinfo  {journal} {Phys. Rev. Lett.}\ }\textbf {\bibinfo {volume} {48}},\ \bibinfo {pages} {73} (\bibinfo {year} {1982})}\BibitemShut {NoStop}%
\bibitem [{\citenamefont {Du}\ \emph {et~al.}(2020)\citenamefont {Du}, \citenamefont {Baru}, \citenamefont {Guo}, \citenamefont {Hanhart}, \citenamefont {Mei\ss{}ner}, \citenamefont {Nefediev},\ and\ \citenamefont {Strakovsky}}]{Du:2020bqj}%
  \BibitemOpen
  \bibfield  {author} {\bibinfo {author} {\bibfnamefont {M.-L.}\ \bibnamefont {Du}}, \bibinfo {author} {\bibfnamefont {V.}~\bibnamefont {Baru}}, \bibinfo {author} {\bibfnamefont {F.-K.}\ \bibnamefont {Guo}}, \bibinfo {author} {\bibfnamefont {C.}~\bibnamefont {Hanhart}}, \bibinfo {author} {\bibfnamefont {U.-G.}\ \bibnamefont {Mei\ss{}ner}}, \bibinfo {author} {\bibfnamefont {A.}~\bibnamefont {Nefediev}}, \ and\ \bibinfo {author} {\bibfnamefont {I.}~\bibnamefont {Strakovsky}},\ }\href {\doibase 10.1140/epjc/s10052-020-08620-5} {\bibfield  {journal} {\bibinfo  {journal} {Eur. Phys. J. C}\ }\textbf {\bibinfo {volume} {80}},\ \bibinfo {pages} {1053} (\bibinfo {year} {2020})},\ \Eprint {http://arxiv.org/abs/2009.08345} {arXiv:2009.08345 [hep-ph]} \BibitemShut {NoStop}%
\bibitem [{\citenamefont {Guo}\ \emph {et~al.}(2023)\citenamefont {Guo}, \citenamefont {Ji}, \citenamefont {Liu},\ and\ \citenamefont {Yang}}]{Guo:2023pqw}%
  \BibitemOpen
  \bibfield  {author} {\bibinfo {author} {\bibfnamefont {Y.}~\bibnamefont {Guo}}, \bibinfo {author} {\bibfnamefont {X.}~\bibnamefont {Ji}}, \bibinfo {author} {\bibfnamefont {Y.}~\bibnamefont {Liu}}, \ and\ \bibinfo {author} {\bibfnamefont {J.}~\bibnamefont {Yang}},\ }\href {\doibase 10.1103/PhysRevD.108.034003} {\bibfield  {journal} {\bibinfo  {journal} {Phys. Rev. D}\ }\textbf {\bibinfo {volume} {108}},\ \bibinfo {pages} {034003} (\bibinfo {year} {2023})},\ \Eprint {http://arxiv.org/abs/2305.06992} {arXiv:2305.06992 [hep-ph]} \BibitemShut {NoStop}%
\bibitem [{\citenamefont {Hackett}\ \emph {et~al.}(2024)\citenamefont {Hackett}, \citenamefont {Pefkou},\ and\ \citenamefont {Shanahan}}]{Hackett:2023rif}%
  \BibitemOpen
  \bibfield  {author} {\bibinfo {author} {\bibfnamefont {D.~C.}\ \bibnamefont {Hackett}}, \bibinfo {author} {\bibfnamefont {D.~A.}\ \bibnamefont {Pefkou}}, \ and\ \bibinfo {author} {\bibfnamefont {P.~E.}\ \bibnamefont {Shanahan}},\ }\href {\doibase 10.1103/PhysRevLett.132.251904} {\bibfield  {journal} {\bibinfo  {journal} {Phys. Rev. Lett.}\ }\textbf {\bibinfo {volume} {132}},\ \bibinfo {pages} {251904} (\bibinfo {year} {2024})},\ \Eprint {http://arxiv.org/abs/2310.08484} {arXiv:2310.08484 [hep-lat]} \BibitemShut {NoStop}%
\bibitem [{\citenamefont {Cao}\ \emph {et~al.}(2025{\natexlab{a}})\citenamefont {Cao}, \citenamefont {Guo}, \citenamefont {Li},\ and\ \citenamefont {Yao}}]{Cao:2024zlf}%
  \BibitemOpen
  \bibfield  {author} {\bibinfo {author} {\bibfnamefont {X.-H.}\ \bibnamefont {Cao}}, \bibinfo {author} {\bibfnamefont {F.-K.}\ \bibnamefont {Guo}}, \bibinfo {author} {\bibfnamefont {Q.-Z.}\ \bibnamefont {Li}}, \ and\ \bibinfo {author} {\bibfnamefont {D.-L.}\ \bibnamefont {Yao}},\ }\href {\doibase 10.1038/s41467-025-62278-9} {\bibfield  {journal} {\bibinfo  {journal} {Nature Commun.}\ }\textbf {\bibinfo {volume} {16}},\ \bibinfo {pages} {6979} (\bibinfo {year} {2025}{\natexlab{a}})},\ \Eprint {http://arxiv.org/abs/2411.13398} {arXiv:2411.13398 [hep-ph]} \BibitemShut {NoStop}%
\bibitem [{\citenamefont {Cao}\ \emph {et~al.}(2025{\natexlab{b}})\citenamefont {Cao}, \citenamefont {Guo}, \citenamefont {Li}, \citenamefont {Wu},\ and\ \citenamefont {Yao}}]{Cao:2025dkv}%
  \BibitemOpen
  \bibfield  {author} {\bibinfo {author} {\bibfnamefont {X.-H.}\ \bibnamefont {Cao}}, \bibinfo {author} {\bibfnamefont {F.-K.}\ \bibnamefont {Guo}}, \bibinfo {author} {\bibfnamefont {Q.-Z.}\ \bibnamefont {Li}}, \bibinfo {author} {\bibfnamefont {B.-W.}\ \bibnamefont {Wu}}, \ and\ \bibinfo {author} {\bibfnamefont {D.-L.}\ \bibnamefont {Yao}},\ }\href@noop {} {\  (\bibinfo {year} {2025}{\natexlab{b}})},\ \Eprint {http://arxiv.org/abs/2507.05375} {arXiv:2507.05375 [hep-ph]} \BibitemShut {NoStop}%
\bibitem [{\citenamefont {Cao}\ \emph {et~al.}(2020)\citenamefont {Cao}, \citenamefont {Guo}, \citenamefont {Liang}, \citenamefont {Wu}, \citenamefont {Xie}, \citenamefont {Xie}, \citenamefont {Yang},\ and\ \citenamefont {Zou}}]{Cao:2019gqo}%
  \BibitemOpen
  \bibfield  {author} {\bibinfo {author} {\bibfnamefont {X.}~\bibnamefont {Cao}}, \bibinfo {author} {\bibfnamefont {F.-K.}\ \bibnamefont {Guo}}, \bibinfo {author} {\bibfnamefont {Y.-T.}\ \bibnamefont {Liang}}, \bibinfo {author} {\bibfnamefont {J.-J.}\ \bibnamefont {Wu}}, \bibinfo {author} {\bibfnamefont {J.-J.}\ \bibnamefont {Xie}}, \bibinfo {author} {\bibfnamefont {Y.-P.}\ \bibnamefont {Xie}}, \bibinfo {author} {\bibfnamefont {Z.}~\bibnamefont {Yang}}, \ and\ \bibinfo {author} {\bibfnamefont {B.-S.}\ \bibnamefont {Zou}},\ }\href {\doibase 10.1103/PhysRevD.101.074010} {\bibfield  {journal} {\bibinfo  {journal} {Phys. Rev. D}\ }\textbf {\bibinfo {volume} {101}},\ \bibinfo {pages} {074010} (\bibinfo {year} {2020})},\ \Eprint {http://arxiv.org/abs/1912.12054} {arXiv:1912.12054 [hep-ph]} \BibitemShut {NoStop}%
\bibitem [{\citenamefont {Dutrieux}\ \emph {et~al.}(2021)\citenamefont {Dutrieux}, \citenamefont {Lorc\'e}, \citenamefont {Moutarde}, \citenamefont {Sznajder}, \citenamefont {Trawi\'nski},\ and\ \citenamefont {Wagner}}]{Dutrieux:2021nlz}%
  \BibitemOpen
  \bibfield  {author} {\bibinfo {author} {\bibfnamefont {H.}~\bibnamefont {Dutrieux}}, \bibinfo {author} {\bibfnamefont {C.}~\bibnamefont {Lorc\'e}}, \bibinfo {author} {\bibfnamefont {H.}~\bibnamefont {Moutarde}}, \bibinfo {author} {\bibfnamefont {P.}~\bibnamefont {Sznajder}}, \bibinfo {author} {\bibfnamefont {A.}~\bibnamefont {Trawi\'nski}}, \ and\ \bibinfo {author} {\bibfnamefont {J.}~\bibnamefont {Wagner}},\ }\href {\doibase 10.1140/epjc/s10052-021-09069-w} {\bibfield  {journal} {\bibinfo  {journal} {Eur. Phys. J. C}\ }\textbf {\bibinfo {volume} {81}},\ \bibinfo {pages} {300} (\bibinfo {year} {2021})},\ \Eprint {http://arxiv.org/abs/2101.03855} {arXiv:2101.03855 [hep-ph]} \BibitemShut {NoStop}%
\bibitem [{\citenamefont {Cao}\ and\ \citenamefont {Zhang}(2023)}]{Cao:2023wyz}%
  \BibitemOpen
  \bibfield  {author} {\bibinfo {author} {\bibfnamefont {X.}~\bibnamefont {Cao}}\ and\ \bibinfo {author} {\bibfnamefont {J.}~\bibnamefont {Zhang}},\ }\href {\doibase 10.1140/epjc/s10052-023-11663-z} {\bibfield  {journal} {\bibinfo  {journal} {Eur. Phys. J. C}\ }\textbf {\bibinfo {volume} {83}},\ \bibinfo {pages} {505} (\bibinfo {year} {2023})},\ \Eprint {http://arxiv.org/abs/2301.06940} {arXiv:2301.06940 [hep-ph]} \BibitemShut {NoStop}%
\bibitem [{\citenamefont {Graczyk}\ and\ \citenamefont {Juszczak}(2014)}]{Graczyk:2014lba}%
  \BibitemOpen
  \bibfield  {author} {\bibinfo {author} {\bibfnamefont {K.~M.}\ \bibnamefont {Graczyk}}\ and\ \bibinfo {author} {\bibfnamefont {C.}~\bibnamefont {Juszczak}},\ }\href {\doibase 10.1103/PhysRevC.90.054334} {\bibfield  {journal} {\bibinfo  {journal} {Phys. Rev. C}\ }\textbf {\bibinfo {volume} {90}},\ \bibinfo {pages} {054334} (\bibinfo {year} {2014})},\ \Eprint {http://arxiv.org/abs/1408.0150} {arXiv:1408.0150 [hep-ph]} \BibitemShut {NoStop}%
\bibitem [{\citenamefont {Cui}\ \emph {et~al.}(2021)\citenamefont {Cui}, \citenamefont {Binosi}, \citenamefont {Roberts},\ and\ \citenamefont {Schmidt}}]{Cui:2021vgm}%
  \BibitemOpen
  \bibfield  {author} {\bibinfo {author} {\bibfnamefont {Z.-F.}\ \bibnamefont {Cui}}, \bibinfo {author} {\bibfnamefont {D.}~\bibnamefont {Binosi}}, \bibinfo {author} {\bibfnamefont {C.~D.}\ \bibnamefont {Roberts}}, \ and\ \bibinfo {author} {\bibfnamefont {S.~M.}\ \bibnamefont {Schmidt}},\ }\href {\doibase 10.1103/PhysRevLett.127.092001} {\bibfield  {journal} {\bibinfo  {journal} {Phys. Rev. Lett.}\ }\textbf {\bibinfo {volume} {127}},\ \bibinfo {pages} {092001} (\bibinfo {year} {2021})},\ \Eprint {http://arxiv.org/abs/2102.01180} {arXiv:2102.01180 [hep-ph]} \BibitemShut {NoStop}%
\bibitem [{\citenamefont {Ye}\ \emph {et~al.}(2018)\citenamefont {Ye}, \citenamefont {Arrington}, \citenamefont {Hill},\ and\ \citenamefont {Lee}}]{Ye:2017gyb}%
  \BibitemOpen
  \bibfield  {author} {\bibinfo {author} {\bibfnamefont {Z.}~\bibnamefont {Ye}}, \bibinfo {author} {\bibfnamefont {J.}~\bibnamefont {Arrington}}, \bibinfo {author} {\bibfnamefont {R.~J.}\ \bibnamefont {Hill}}, \ and\ \bibinfo {author} {\bibfnamefont {G.}~\bibnamefont {Lee}},\ }\href {\doibase 10.1016/j.physletb.2017.11.023} {\bibfield  {journal} {\bibinfo  {journal} {Phys. Lett. B}\ }\textbf {\bibinfo {volume} {777}},\ \bibinfo {pages} {8} (\bibinfo {year} {2018})},\ \Eprint {http://arxiv.org/abs/1707.09063} {arXiv:1707.09063 [nucl-ex]} \BibitemShut {NoStop}%
\end{thebibliography}%

\end{document}